\documentclass{webofc}
\usepackage[varg]{txfonts}   
\usepackage{braket}
\begin{document}
\title{Self-consistent energy density functional approaches to the crust of neutron stars}
%
%

\author{\firstname{Takashi} \lastname{Nakatsukasa}\inst{1,2}
}

\institute{Center for Computational Sciences and Faculty of Pure and Applied Sciences, University of Tsukuba Tsukuba 305-8577, Japan
\and
           RIKEN Nishina Center, Wako 351-0198, Japan 
          }

\abstract{%
In order to study structure of the crust in neutron stars,
we develop a finite-temperature Skyrme-Hartree-Fock method
in the full three-dimensional coordinate space
using the Fermion operator expansion method.
It provides us with a possible order-$N$ approach to
non-uniform neutron-star matters.
}
\maketitle
\section{Introduction}
\label{intro}
The neutron star is a compact star whose radius is about 10 km,
which can be regarded as a giant nucleus with macroscopic numbers
of hadrons (baryons and mesons).
The core part, the central region of the neutron star,
is supposed to be uniform nuclear (hadron) matter.
However, moving toward the surface,
it changes into non-uniform nuclear matter composed of
neutron-rich nuclei and neutron matter,
called ``inner crust''.
In the vicinity of the boundary between the core and the inner crust,
exotic phases, called ``nuclear pasta'' are predicted.
Since a new exotic structure may appear in certain conditions,
the calculation without any symmetry restriction is desired.

Structure of these non-uniform exotic nuclear matter have been
studied with the Thomas-Fermi (local density) approximation \cite{Oya93}
and/or with the Wigner-Seitz approximation \cite{NV73}.
Although these two approximations are complimentary to each other,
neither is able to take proper account of transport properties in the
non-uniform nuclear matter.
Thus, it is highly desirable to go beyond these approximations.

Recently, the three-dimensional (3D) coordinate-space solver of the
Hartree-Fock-Bogoliubov (HFB) method has been developed \cite{JBRW17}.
In order to abstain from diagonalizing
the large HFB Hamiltonian matrix, which is computationally demanding,
the contour integration in the complex energy plane together with
iterative solutions of
the shifted linear algebraic equations is adopted
to construct densities.
The method has been extended to the finite-temperature HFB
method, by including the Matsubara frequencies as the
imaginary shifts \cite{KN20}.
The method was successfully applied to non-uniform nuclear matter
with superfluid neutrons at finite temperature.

In this paper, we present another coordinate-space solver
using the Fermion operator expansion (FOE) method \cite{GT95}.
The method was originally developed in the computational
condensed matter physics,
and known as one of order-$N$ ($O(N)$) methods \cite{WJ02}.
The conventional approaches to the density functional theory scales
as $N^3$, where $N$ is the number of particles.
This $N^3$-scaling sets a severe limitation to the system size.
In contrast, the $O(N)$ method scales as $N^1$ in ideal cases.
The foundation of the $O(N)$ method is provided by
``principle of nearsightedness'' \cite{WJ02},
namely a consequence of quantum mechanical decoherence
among many particles.
Calculation of the non-uniform neutron star matter may require
us to treat more than thousands of nucleons (baryons)
in a large space.
In this paper, we demonstrate a test calculation of the FOE method
for non-uniform nuclear matter at finite temperature.

\section{Fermion operator expansion method}
\label{sec:FOE}
Let us briefly recapitulate the FOE method.
$\hat{H}$ is the one-body mean-field Hamiltonian,
satisfying $\hat{H}\ket{n}=E_n\ket{n}$.
The one-body density at a temperature $T=1/\beta$ is given by
\begin{equation}
	\hat{\rho}_T=\sum_n f(E_n)\ket{n}\bra{n}
	=f(\hat{H})\sum_n \ket{n}\bra{n}
	=f(\hat{H}) ,
\end{equation}
where $f(x)$ is the Fermi-Dirac distribution function,
$f(x)=(1+e^{\beta(x-\mu)})^{-1}$.
Next, the Fermi-Dirac function
is approximated by a series of polynomial functions $\{ T_j(x) \}$;
$f(x)=\sum_j a_j T_j(x)$.
Then, the density, represented by $f(\hat{H})$, can be written as
\begin{equation}
	\hat{\rho}_T \approx \sum_{j=0}^M a_j T_j(\hat{H}) ,
\end{equation}
where $T_j(x)$ is a polynomial of the $j$-th degree.
Here, the summation is truncated at the maximum degree $M$.
In order to obtain a reasonable description of the
Fermi-Dirac function,
roughly speaking, $M$ should increase at lower $T$.
The density can be constructed by
the $M$ times operation of the Hamiltonian.
We adopt the Chebyshev polynomials following ref.~\cite{GT95}.
They can be determined by the following recursion relations:
\begin{equation}
	T_j(x)=2x T_{j-1}(x)-T_{j-2}(x) ,
\end{equation}
with $T_0(x)=1$ and $T_1(x)=x$.
Therefore, the operation of the polynomial of the Hamiltonian
on a state $\ket{\psi}$,
$\ket{\psi^{(j)}}\equiv T_j(\hat{H})\ket{\psi}$,
is calculated as
\begin{equation}
	\ket{\psi^{(j)}}=2\hat{H}\ket{\psi^{(j-1)}} -\ket{\psi^{(j-2)}} ,
\end{equation}
with $\ket{\psi^{(0)}}=\ket{\psi}$ and $\ket{\psi^{(1)}}=\hat{H}\ket{\psi}$.
The density is given by
\begin{equation}
	\hat{\rho}_T\ket{\psi} = \sum_{j=0}^M a_j \ket{\psi^{(j)}} .
	\label{rho_T}
\end{equation}
We need to perform the calculation for, at least, $N$ different states,
$\{ \ket{\psi_k},k=1,\cdots,N\}$.
If the Hamiltonian is sparse,
the total computational cost is estimated as $O(N^2)$.
However, when the ``nearsightedness'' exists,
it can be reduced to $O(N)$ \cite{WJ02}.

\section{Application of the FOE method}
\label{sec:BKN}

In this paper, we apply the FOE method to the BKN energy density
functional \cite{BKN76}
which is a simplified Skyrme-like functional without
the spin-orbit coupling and assumes the identical densities of
protons and neutrons.
For simplicity, we neglect the pairing correlation, thus,
it is the finite-temperature Hartree-Fock calculation,
instead of finite-temperature HFB.
The calculation is performed using the 3D coordinate-space representation
discretized in the mesh \cite{NIY07}.
The number of mesh points is $25^3$ with the cubic mesh of (1 fm)$^3$.
The baryon density and the proton ratio
are fixed at $\rho_B=0.03$ fm$^{-3}$
and at $Y_p=0.5$ (symmetric nuclear matter), respectively.

\begin{figure}[t]
\centering
\includegraphics[width=12cm,clip]{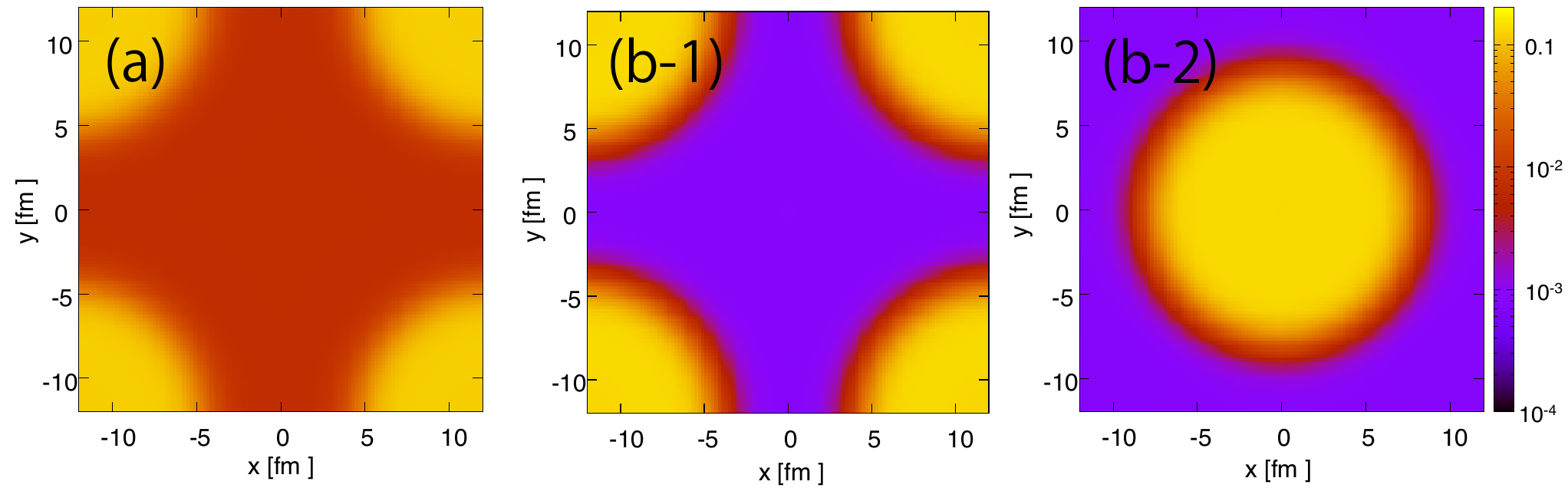}
	\vspace{-10pt}
\caption{Calculated nucleonic density distribution
	(log scale) of the bcc phase in the $xy$ plane
	at $\rho_B=0.03$ fm$^{-3}$ in the planes
	$z=|z_{\rm max}|$ fm (panels (a) and (b-1)),
	$z=0$ (b-2).
	The temperature is $T=5$ MeV (b) and $T=10$ MeV (a).
	}
\label{fig:bcc}       
	\vspace{-10pt}
\end{figure}

We first set the temperature $T=5$ MeV,
then, perform the self-consistent iterative calculation starting from
the body-centered configuration (bcc) as the initial state.
The converged state also has a bcc-like structure:
In the ``sea'' of symmetric nuclear matter whose density is
about $8.3\times 10^{-4}$ fm$^{-3}$,
a ``nucleus'' (excess density) is embedded at the center and 1/8 of a nucleus
is at each corner of the cubic box (Fig.~\ref{fig:bcc}(b)).
In total, there are two nuclei in the adopted space 
of volume $V=25^3$ fm$^3$.
The number of nucleons in the space is $\rho_B\times V=468.75$,
so that, at every iteration,
the chemical potential $\mu$ is adjusted to produce
this particle number.
Increasing the temperature (Fig.~\ref{fig:bcc}(a)), 
the nucleons are gradually leaking from ``nucleus'' to the ``sea'',
approaching to the uniform matter.
It should be noted that, since the symmetric matter is assumed,
the result is not realistic for neutron star inner crust.
However, it is still meaningful for testing performance
of the FOE method applied to nuclear matter.

The FOE method is suitable for parallel computing, because
calculation of eq. (\ref{rho_T}) can be performed independently
for different $\ket{\psi}$.
The FOE method is extremely efficient at high temperature,
since the Fermi-Dirac function becomes smooth and
is well approximated by a small number of polynomials.
The computational cost is roughly proportional to $M$
in eq. (\ref{rho_T}).
$M$ is order of thousands at $T\sim 100$ keV,
while it is order ten at $T\sim 10$ MeV.

This work is supported by JSPS KAKENHI Grant No.18H01209
and No.19H05142.
This research used computational resources provided by
Joint Center for Advanced High Performance Computing (JCAHPC)
through the HPCI System Research Project (Project ID: hp200069)
and through Multidisciplinary Cooperative Research Program
in Center for Computational Sciences, University of Tsukuba.





%
\bibliography{current,myself,nuclear_physics}
%
%
%
%

\end{document}